\documentclass[review,12pt]{article}
\usepackage{natbib}
\usepackage[table,dvipsnames]{xcolor}
\usepackage{authblk}
\usepackage{dsfont}
\usepackage[T1]{fontenc}
\usepackage{hyperref}
\usepackage{orcidlink}

\begin{document}


\title{Finding Early Adopters of Innovation in Social Networks}

\date{}

\author[1,2]{Balázs R. Sziklai
\texorpdfstring{%
\orcidlink{0000-0002-0068-8920}
}{}}
\affil[1]{%
Centre for Economic and Regional Studies, Budapest, Hungary}
\affil[2]{%
Department of Operations Research and Actuarial Sciences, Corvinus University of Budapest}

\author[1,3]{Balázs Lengyel
\texorpdfstring{%
\orcidlink{0000-0001-5196-5599}
}{}}
\affil[3]{%
Corvinus Institute for Advanced Studies, Corvinus University of Budapest}

\maketitle

\begin{abstract}Social networks play a fundamental role in the diffusion of innovation through peers' influence on adoption. Thus, network position including a wide range of network centrality measures have been used to describe individuals' affinity to adopt an innovation and their ability to propagate diffusion. Yet, social networks are assortative in terms of susceptibility and influence and in terms of network centralities as well.
This makes the identification of influencers difficult especially since susceptibility and centrality does not always go hand in hand.
Here we propose the Top Candidate algorithm, an expert recommendation method, to rank individuals based on their perceived expertise, which resonates well with the assortative nature of innovators and early adopters.
Leveraging adoption data from two online social networks that are assortative in terms of adoption but represent different levels of assortativity of network centralities, we demonstrate that the Top Candidate ranking is more efficient in capturing early adopters than other widely used indices. Top Candidate nodes adopt earlier and have higher reach among innovators, early adopters and early majority than nodes highlighted by other methods. These results suggest that the Top Candidate method can identify good seeds for influence maximization campaigns on social networks.
\end{abstract}




\section{Introduction}

Most individuals adopt an innovation by imitating their influential peers \citep{Rogers1962,bass1969new} that underlines the role of social networks in the diffusion of new products, technologies or ideas \citep{Granovetter1978, valente1996social}. Network scientists argue that the structure of social networks can explain the underlying mechanisms of social influence and adoption: highly connected nodes have more influence than others \citep{pastor2015epidemic}, while diffusion is more likely in tightly connected cliques and less likely across them \citep{centola2007complex}. \cite{Wang2019} paint a more nuanced picture as they found that although simple messages spread effectively via network hubs, for complex stories, the influence of 'ordinary people' (individuals that are less connected) is more important.

A central part of this discussion has led to the "influence maximization" problem (IM) \citep{Kempe2003}, which aims to identify the ideal seed nodes that a marketing campaign should target to achieve maximum impact, given pre-defined diffusion models. The IM is NP-hard; thus, many use heuristics to find the seed nodes and start optimization by assuming that nodes with high network centrality (e.g.\ degree) are influential spreaders \citep{Kitsak2010, de2014role} and run diffusion simulations, most notably using the linear threshold and the independent cascade models. However, these models fail to  capture an important feature that is observed in real life networks: homophily, the tendency that similar individuals are more likely to be connected than dissimilar ones.

Homophily, also referred to as assortativity in relation with social networks \citep{newman2002assortative}, is a general phenomenon  \citep{McPherson2001,Cho2012} that has a fundamental role in innovation spreading \citep{Anwar2021}. Ignoring this effect poses a major problem for seed indentification in influence maximization when the sole source of information is the network structure \citep{aral2018social}. 
Social influence and centrality are difficult to disentangle without knowing at least some of the early adopters of the specific innovation \citep{banerjee2013diffusion, toole2012modeling}. There are plenty of reasons why a central individual may be reluctant to participate in a campaign or may not be susceptible to the marketing message. The most prominent one is risk-averseness. Subscribing to a new trend or technology needs commitment and entails social risk -- not everyone is willing to do that. Central agents with many friends may particularly feel the social pressure to be conformist and to avoid eccentric behavior. Innovators and early adopters, on the other hand, are known to possess psychological traits that makes them perfect subject for the early market of an innovation.

In this paper, we aim to contribute to this discussion in two ways. First, empirical data on adoption dynamics from two online social networks enable us to investigate how network structure can be useful to identify innovators and early adopters in innovation diffusion. Second, we propose a ranking of the users based on the so-called Top Candidate method \citep{Sziklai2018} -- an expert selection algorithm  that exhibits features resembling assortativity. We argue that (i) these contributions together provide new heuristics for seed selection in influence maximization (ii) any framework that neglects assortativity of social networks when testing heuristics is prone to produce inaccurate predictions.


We compare the Top Candidate ranking with seven well-known centrality measures on two online social networks: iWiW from Hungary and Pokec from Slovakia. Registration days of users are known in both networks, both are assortative in terms of adoption time but represent different levels of assortativity in network centralities. We look at the top 1000 nodes of the Top Candidate ranking and the other seven alternative measures and plot how the date of registration is distributed over time.

We find that the Top Candidate ranking is more efficient in capturing innovators and early adopters than other widely used indicators. Top Candidate nodes adopt earlier and have higher reach among innovators, early adopters and early majority than nodes highlighted by other methods. These results suggest that the Top Candidate method can identify good seeds for influence maximization campaigns on social networks.

\section{Literature overview}\label{sec:literature}


\subsection{Early adopters as experts}\label{sec:characteristics}
The identification of innovators and early adopters is key for marketing campaigns and their characterization received considerable attention.
The literature converges toward the conclusion that innovators and early adopters stand out from their peers.

\cite{Rogers2003} describes innovators as venturesome individuals who can cope with a high degree of uncertainty, and early adopters as a group with high socio-economic status. 
\cite{Moore2014} depicts innovators as technology enthusiasts, or geeks and early adopters as visionaries who are willing to take high risk.

A field study by \cite{Brancheau1990} supports hypotheses that early
adopters  were  more highly educated, more attuned to mass media, more involved in interpersonal communication, and more likely to be opinion leaders. \cite{Eastlick1999} reports that social risk negatively relates to the tendency to be a potential innovator and potential innovators possessed significantly stronger opinion leadership. A dutch survey shows that early adopters are likely to be highly mobile, have a high socio-economic status, high levels of education and high personal incomes \citep{Zijlstra2020}. Finally, \cite{Muller2006} provides empirical evidence that the average time at which the main market outnumbers the early market is indeed when 16\% of the market has already adopted the product -- giving support \cite{Rogers1962}'s somewhat arbitrary (or rather inspired) division of adopter sets.



Another important concept is market mavenness \citep{Feick1987}. Market mavens are consumers who are highly involved in a market. They have information about many kinds of products and shops, and they enjoy sharing their knowledge. Peers often seek out their opinion and rely on their expertise. \cite{Goldsmith2003} finds that consumer innovativeness and market mavenism positively correlates, although they argue that market mavens and innovators are distinct groups.

Directly related to the context of this study, \cite{Lynn2017} explores the relationship between personality traits of early adopters of social network sites. They report that extraversion, openness and conscientiousness impact positively and significantly on information sharing, and negatively on rumor sharing. On the other hand both, information sharing and rumor sharing impact positively and significantly on the centrality of early adopters. The seemingly contradictory observations, can be explained away by separating the social status of opinion leadership and the influencing capacity of the agent which relates more to network centrality.

To sum up, innovators and early adopters stand out in their personal characteristics. Thus, marketing campaigns have usually targeted and labelled them as experts to convince society. However, both \cite{Lynn2017} and \cite{Dedehayir2017}  argue that a distinction has to be made between opinion leadership and innovativeness. Even \cite{Rogers2003} affirms that opinion leaders are not necessarily innovators.

\subsection{Early Adoption and homophily in network diffusion}\label{sec:diffusion}
In the Influence Maximization framework \citep{Kempe2003}, few papers addressed other node characteristics concentrating in network communities that can help to predict the future popularity of novelty. For example, influential individuals can form clusters that can help the early propagation of an idea \citep{Aral2012}. \cite{Weng2014} build a predictive model for meme popularity using three classes of features: network topology, community diversity, and growth rate. They found that community related features are the most powerful predictors of future success.
\cite{Hajdu2020} study the community structure of public transportation networks and finds that transmission probabilities depend on the community structure.
\cite{Calio2021} study attribute based seed diversification. They argue that a seed set with different characteristics (age, gender, etc.) might be more successful in information-propagation. \cite{Rahimkhani2015} identifies the community structures of the input graph then chooses a number of representative nodes to form the final output of the proposed algorithm.

However, this literature has largely overlooked a phenomenon inherent is social networks and diffusion dynamics alike: the role of homophily \citep{McPherson2001}. It has long been recognized that a behavior can spread in society only when those most prone to it are surrounded by peers who are somewhat less but almost equally open to it's adoption \citep{Granovetter1978}. In other words, innovators must be connected to early adopters such that adoption can penetrate in their communities and later influence the rest of the market too, otherwise the innovation will not spread \citep{watts2002simple}. Adoption dynamics can be predicted at small scales only by assuming homophily of adoption \citep{toole2012modeling}. Despite the importance of adoption homophily in networks, it has been largely ignored in influence maximization modeling \citep{aral2018social}.

Instead, a usual assumption to find the seed nodes for Influence Maximization is that network structure alone can quantify influence. For example, nodes with high network centrality (e.g.\ degree) are usually considered as influential spreaders \citep{Kitsak2010, de2014role}.

Finally, the presence of assortativity implies that not every connection is equally important in the diffusion. However, the literature also ignored the problem of determining where the probabilities of influence between users comes from \citep{Goyal2010}. Recently, \cite{Qiang2019} proposed two learning models that are aimed at understanding person-to-person influence in information diffusion from historical cascades, while \cite{Bota2015} and \cite{Bota2016} considered the Inverse Infection Problem as a way to estimate the hidden edge infection probabilities.





\section{Data and methodology}\label{sec:methodology}

In this paper, we propose the Top Candidate method that can identify innovators and early adopters in social networks more efficiently than other widely used network centrality measures, by using network structure as the only source of information. We compare the ranking induced by the Top Candidate method with seven other centrality measures by using data from two online social networks.

\subsection{Data}\label{sec:data}

Our empirical analysis leverages data retrieved from two social media platforms. The first platform is called iWiW (international who is who) that was an early Hungarian version of online social networks aiming to link pre-existing friends and an outstanding online innovation of its' time. The iWiW platform existed between 2002 and 2014. It was the most visited website in the country in the mid 2000s, but failed the competition with Facebook that started in Hungary from 2008. Pokec is a still functioning Slovakian dating and chatting website with a purpose of meeting new people.

These data sources provide unique opportunity to understand how network structure can help us identifying early adopters of an innovation. Both data sources contain the date of individual registration to the websites that is used a proxy of adoption. Data also includes the identifiers of friends that enables us to generate social networks. The iWiW dataset has been used in previous work to describe and model the innovation diffusion process \citep{torok2017cascading,lengyel2020role,bokanyi2022urban}.

Here, we use a 10\% sample of the iWiW data that contains 271 913 nodes 2 712 587 edges. The Pokec network contains 277 695 nodes and 2 122 778 edges. Access to iWiW data was provided to us by a non-disclosure agreement with the data owner company. Pokec data are open access at \url{https://snap.stanford.edu/data/soc-pokec.html}.

\subsection{The Top Candidate method}\label{sec:top_cand}
\textbf{Top candidate} (TC) algorithm is a group identification method designed to find experts on recommendation networks \citep{Sziklai2018,Sziklai2021a}. The algorithm takes a network as an input and outputs a list of experts. With a parameter, $\alpha\in [0,1]$ we can adjust how exclusive our list should be. Each agent nominates $\alpha$ fraction of their most popular neighbours as experts, where popularity is based on (weighted) in-degree. In the beginning, every agent is labeled as an expert, then in successive rounds we remove the nominations of agents who were not nominated by anyone until we obtain a stable set. The underlying idea resembles homophily: experts identify other experts much more effectively than amateurs. Thus, in the set of experts (i) each expert should be nominated by another expert and (ii) each nominee of an expert should be also included in the set -- this property is called \emph{stability}.

One advantage of the Top Candidate algorithm is its axiomatic characterization. It is the unique method\footnote{In the Social Choice literature a group identification method, that takes a recommendation network as input and outputs a list of nodes, is called a \emph{collective identity function}. CIFs are special kinds of centrality measures that map the node set into $\{0,1\}^n$ instead of the usual $\mathds{R}^n$.} that satisfies stability, exhaustiveness and decisiveness. Exhaustiveness ensures that all possible experts are recognized on the network, not just a subset, and decisiveness guarantees that at least one expert is selected if reasonable choices are presented. There are other centralities that feature characterizations, most notably PageRank \citep{Was2018b}, Generalized Degree \citep{Csato2017} and the Shapley-value \citep{Shapley1953,Young1985}, but it is less clear how these relate to sociodemographic properties of the nodes.

\subsection{Network centralities}\label{sec:centralities}

\noindent We compute seven other measures on the data to asses their ability in finding innovators and early adopters.

\textbf{Degree} represents the number of connections that a user has. It is a natural benchmark for the user's centrality. Another classical measure is \textbf{Harmonic centrality}. It is a distance-based measure proposed by \citet{Marchiori2000}. Harmonic centrality of a node, $\mathbf{u}$ is the sum the reciprocal of distances between $\mathbf{u}$ and every other node in the network. Disconnected node pairs have infinite distance, thus the reciprocal is defined as zero. Peripheral agents, who are many handshakes away from most of the other users, thus have a small Harmonic centrality.

\textbf{PageRank} (PR), introduced by \citet{Page99}, is a close relative of Eigenvector centrality \citep{Bonacich1972}. The latter assigns centrality scores to nodes based on the eigenvector of the adjacency matrix of the underlying graph. The method breaks down if the graph is not strongly connected. PageRank rectifies this by (i) connecting sink nodes (\emph{i.e.}\ nodes with no leaving arc) with every other node through a link and (ii) redistributing some value uniformly among the nodes. Redistribution is parameterised by the so called damping factor, $\alpha\in(0,1)$. PageRank was designed to model a random walk on the World Wide Web. We start from an arbitrary webpage. On any subsequent step, we leave the current webpage with equal probability on one of the departing links. After each step, we have a $(1-\alpha)$ probability to restart the walk at a random node. The probability that we occupy node $\mathbf{u}$ as the number of steps tends to infinity is the PageRank value of node $\mathbf{u}$. PageRank composes the core of Google's search engine, but the algorithm is used in a wide variety of applications. The damping value is usually chosen from the interval $(0.7,0.9)$, here we opted for $\alpha=0.8$.

\textbf{Generalised Degree Discount} (GDD) introduced by \citet{Wang2016} was developed specifically for the \emph{independent cascade} network diffusion model.  In this model each active node has a single chance to infect its neighbours, transmission occurring with the probability specified by the arc weights. GDD is a suggested improvement on \emph{Degree Discount} \citep{Chen2009} which constructs a seed group of size $q$ starting from the empty set and adding nodes one by one using a simple heuristic. It primarily looks at the degree of the nodes but also considers how many of their neighbours are already in the seed set. GDD improves this by also taking into account how many of the neighbours' neighbours are spreaders. The spreading parameter of the algorithm was chosen to be $0.05$.

\textbf{$k$-core}, also referred to as, $k$-shell categorizes nodes into layers \citep{Seidman1983,Kitsak2010}. First, it successively delete nodes with only one neighbours. These are assigned a $k$-core value of 1. Then it deletes nodes with two or less neighbours and labels them with a $k$-core value of 2. The process is continued until every node is classified. For instance, every node of a path or a star graph is assigned a $k$-core value of 1, while nodes of a cycle will have a $k$-core value of 2.

\textbf{Linear Threshold Centrality} (LTC), as the name suggests, was developed for the Linear Threshold diffusion model \citep{Riquelme2018}. Given a network, $G$ with node thresholds and arc weights, LTC of a node $\mathbf{u}$ represents the fraction of nodes that $\mathbf{u}$ and its neighbours would manage to activate as a seed set in the Linear Threshold model. Since the social networks we used in our analysis had no data on friendship intensity, we decided to assign a uniform unit weight to each connection. Node thresholds was defined as $0.7$ times the node degree. That is, a user became activated if 70\% of its friends had been active.

\cite{Suri2008} define a cooperative game on the network and derive node centrality by computing the Shapley-value. In this setting, the Shapley-value of a node is the average marginal contribution that a node generates when the seed set is composed by adding nodes one-by-one and any order of the nodes is equally likely. Every node set is assigned a (characteristic function) value. Marginal contribution of a node $\mathbf{u}$ is just the difference between the value of the node set with and without $\mathbf{u}$. There is more than one way how this can be done. We use the G1 game variant proposed by \cite{Michalak2013} who also gave an efficient algorithm to compute the corresponding \textbf{Shapley(G1)-value}. In G1, the characteristic function value of a node set $C$, is the number of nodes in $C$ plus the number of neighbours of $C$. Under this setting, the Shapley-value of a node $u$ is calculated as the sum of reciprocals, $\frac{1}{1+deg(v)}$, for each $v$ belonging to the extended reach of $u$ (the neighbours of $u$ plus $u$ itself).

\section{Results}\label{sec:results}

\subsection{Homophily of adoption}\label{sec:homo}
Before we delve into the performances of centrality measures let us take a look at the networks themselves. Tables~\ref{tab:gr_inter_iWiW} and \ref{tab:gr_inter_Pokec} explore the interconnectedness of adopter groups. iWiW and Pokec paint a similar picture: typically, there are more connections between subsequent groups in the adoption timeline than between other groups. Innovators are mainly befriended with early adopters, who in turn are mainly connected to early majority and so on.

\begin{table*}[!h]
\small\sf
\caption{Group interconnectedness in iWiW.  An entry of the matrix, shows the portion of links that connects the column group to the row group with respect the the column group's total connections. \label{tab:gr_inter_iWiW}}
\begin{tabular}{llllll}
\hline
               & innovators                   & early adopters               & early majority               & late majority                & laggards                     \\
               \hline
innovators     & \cellcolor[HTML]{FFAF8E}23.4 & \cellcolor[HTML]{FFDA97}7.7  & \cellcolor[HTML]{FFE699}2.5  & \cellcolor[HTML]{FFE699}1.3  & \cellcolor[HTML]{FFE699}1.1  \\
early adopters & \cellcolor[HTML]{FF8585}39.8 & \cellcolor[HTML]{FF8C87}36.4 & \cellcolor[HTML]{FFC492}18.6 & \cellcolor[HTML]{FFCE94}11.7 & \cellcolor[HTML]{FFCD94}10.5 \\
early majority & \cellcolor[HTML]{FFA68C}26.8 & \cellcolor[HTML]{FF8585}38.6 & \cellcolor[HTML]{FF8585}48.4 & \cellcolor[HTML]{FF8585}42.5 & \cellcolor[HTML]{FF8585}37.4 \\
late majority  & \cellcolor[HTML]{FFD596}8.4  & \cellcolor[HTML]{FFC893}14.3 & \cellcolor[HTML]{FFB690}25.2 & \cellcolor[HTML]{FF9488}36.3 & \cellcolor[HTML]{FF8C87}35.2 \\
laggards       & \cellcolor[HTML]{FFE699}1.6  & \cellcolor[HTML]{FFE699}3.0  & \cellcolor[HTML]{FFE198}5.2  & \cellcolor[HTML]{FFD696}8.3  & \cellcolor[HTML]{FFBF91}15.9 \\
\hline
total & 100.0 & 100.0 & 100.0 & 100.0 & 100.0 \\
\hline
\end{tabular}
\end{table*}

\begin{table*}[!h]
\small\sf
\caption{Group interconnectedness in Pokec. An entry of the matrix, shows the portion of links that connects the column group to the row group with respect the the column group's total connections.\label{tab:gr_inter_Pokec}}
\begin{tabular}{llllll}
\hline
               & innovators                   & early adopters               & early majority               & late majority                & laggards                     \\
               \hline
innovators     & \cellcolor[HTML]{FFCA94}12.7 & \cellcolor[HTML]{FFE198}4.5  & \cellcolor[HTML]{FFE699}1.6  & \cellcolor[HTML]{FFE699}0.8  & \cellcolor[HTML]{FFE699}0.7  \\
early adopters & \cellcolor[HTML]{FF8585}36.0 & \cellcolor[HTML]{FFA28B}32.7 & \cellcolor[HTML]{FFCA94}16.2 & \cellcolor[HTML]{FFD997}7.3  & \cellcolor[HTML]{FFE098}3.5  \\
early majority & \cellcolor[HTML]{FF8A86}34.5 & \cellcolor[HTML]{FF8585}45.2 & \cellcolor[HTML]{FF8585}51.7 & \cellcolor[HTML]{FF9B8A}35.7 & \cellcolor[HTML]{FFC292}15.9 \\
late majority  & \cellcolor[HTML]{FFC693}13.9 & \cellcolor[HTML]{FFC893}15.6 & \cellcolor[HTML]{FFB58F}27.4 & \cellcolor[HTML]{FF8585}45.7 & \cellcolor[HTML]{FF8585}40.9 \\
laggards       & \cellcolor[HTML]{FFE699}2.9  & \cellcolor[HTML]{FFE699}1.9  & \cellcolor[HTML]{FFE399}3.1  & \cellcolor[HTML]{FFD295}10.5 & \cellcolor[HTML]{FF8A86}39.1 \\
\hline
total & 100.0 & 100.0 & 100.0 & 100.0 & 100.0 \\
\hline
\end{tabular}
\end{table*}

A number of interesting observations can be made. Firstly, the result reinforces Rogers' classification. It is much more obvious why cascade happens the way it does. Innovators have the biggest impact on early adopters because early adopters are the innovators closest - or at least the most numerous - friends.

Secondly, psychological traits do affect the network structure. Rogers' categorization correlates with risk attitudes, extraversion, openness and a number of other traits. It seems that risk-seeking (extrovert, open-minded, etc.) users prefer the company of other risk-seekers, while risk-averse users are more comfortable with other risk-averse individuals. The results are in line with the findings of \cite{Selfhout2010}.

Thirdly, identifying innovators and early adopters does not seem to be a hopeless task anymore. Clearly, these groups form  clusters on the network. Thus, there can be centralities that are systematically better in recognizing them.

These observations have a rather remarkable implication. Researchers of influence maximization frequently validate their algorithms using simulations with either the linear threshold or the independent cascade diffusion models -- these are the most commonly used configurations by far. A basic flaw in these simulations is that thresholds and diffusion probability are chosen at random either independently of the network structure or only having a crude relationship with it. For instance, in the linear threshold model in every simulation the node thresholds (which signify the tendency for the nodes to adopt an innovation) are generated uniformly at random for each node \citep{Kempe2003}. In the independent cascade model the two most common propagation setup is the weighted cascade and the trivalency models \citep{Jung2012}. In the first, the propagation probability on each edge equals to the reciprocal of the degree of the source node, while in the latter it is chosen randomly from the set $\{0.1,0.01,0.001\}$.

In light of Table~\ref{tab:gr_inter_iWiW} and \ref{tab:gr_inter_Pokec} these assumptions lead to a highly unrealistic threshold/propagation probability distributions. In order to obtain a realistic network configuration the distribution should take into consideration the clustering of the adopter sets. 
For instance, thresholds of nodes that belong to innovators or early adopters should be lower in general than thresholds of other nodes. This could be achieved by choosing the thresholds from an interval.
Disregarding the underlying structure introduces a systemic bias that may be favorable for some influence maximization algorithms while detrimental to others.


\begin{table*}[]
\caption{Assortativity of the iWiW and Pokec networks in terms of network centralities. The assortativity index ranges from -1 to +1. Negative values mean that nodes of similar centrality values are not connected while positive values mean that nodes of similar centrality values are connected. \label{tab:assortativity}}
\begin{center}
\begin{tabular}{lll}
\hline
Network centralities        & iWiW   & Pokec  \\ \hline
Degree                      & 0.04   & -0.158 \\
Harmonic centrality         & 0.306  & 0.137  \\
Page Rank                   & 0.02   & -0.13  \\
Generalized Degree Discount & 0.02   & -0.151 \\
k-shell                     & 0.27   & 0.227  \\
Linear Threshold Centrality & 0.05   & -0.06  \\
Shapley value               & -0.004 & -0.095 \\
Top Candidate               & 0.09   & -0.083 \\ \hline
\end{tabular}
\end{center}
\end{table*}

Although the two online social networks are similar in terms of adoption homophily, the assortativity of these networks are different in terms of the network centrality measures described in Section 3.3. Both networks are assortative in terms of Harmonic centrality and k-shell measures (Table \ref{tab:assortativity}). However, Pokec is disassortative in terms of Degree, Generalized Degree Discount, and PageRank. This means that the identification of innovators and early adopters is carried out on networks in which individuals of similar levels of assumed influence are mixed differently.

\subsection{Identification of innovators and early adopters}
Now we turn to the network centrality indicators and their performances in finding innovators and early adopters. We computed the top 1000 nodes according to eight centrality measures on both iWiW and Pokec. If the 1000th and 1001st node tied under some measure, we discarded nodes of the same centrality value randomly until there were only 1000 nodes in the set.

Table \ref{tab:intersection_iwiw} and  \ref{tab:intersection_pokec} show the overlap between the top 1000 nodes of the centralities that we employed in this paper on the iWiW and Pokec networks. Each centrality genuinely differs from the others, although LTC, GDD and PageRank somewhat overlaps with Degree on both networks. In general, $k$-core, TC and Harmonic contain more nodes that are uniquely represented by those centralities.

\begin{table*}[!h]
\footnotesize\sf
\caption{Overlap in the top 1000 nodes of different centralities on the iWiW network. Measures are ordered according to their distance to Degree. \label{tab:intersection_iwiw}}
\begin{tabular}{lllllllll}
           & k-core                       & Harm.                     & TC                           & Shapley(G1)                 & GDD(0.05)                   & PR                     & LTC(0.7)                    & Deg.                       \\
k-core     & \cellcolor[HTML]{FF8585}1000 & \cellcolor[HTML]{FFE699}13   & \cellcolor[HTML]{FFE499}39   & \cellcolor[HTML]{FFE499}39   & \cellcolor[HTML]{FFE299}54   & \cellcolor[HTML]{FFE399}49   & \cellcolor[HTML]{FFE098}84   & \cellcolor[HTML]{FFE098}75   \\
Harm.   &                              & \cellcolor[HTML]{FF8585}1000 & \cellcolor[HTML]{FFD095}241  & \cellcolor[HTML]{FFCB94}295  & \cellcolor[HTML]{FFCB94}290  & \cellcolor[HTML]{FFC392}373  & \cellcolor[HTML]{FFC192}392  & \cellcolor[HTML]{FFC392}374  \\
TC         &                              &                              & \cellcolor[HTML]{FF8585}1000 & \cellcolor[HTML]{FFC192}394  & \cellcolor[HTML]{FFB890}485  & \cellcolor[HTML]{FFB990}477  & \cellcolor[HTML]{FFBA90}470  & \cellcolor[HTML]{FFB890}482  \\
\multicolumn{2}{l}{Shapley (G1)}          &                              &                              & \cellcolor[HTML]{FF8585}1000 & \cellcolor[HTML]{FFA88D}652  & \cellcolor[HTML]{FF9E8B}753  & \cellcolor[HTML]{FFAB8D}621  & \cellcolor[HTML]{FFAA8D}629  \\
GDD(0.05) &                              &                              &                              &                              & \cellcolor[HTML]{FF8585}1000 & \cellcolor[HTML]{FF9889}813  & \cellcolor[HTML]{FF9A8A}788  & \cellcolor[HTML]{FF9689}828  \\
PR   &                              &                              &                              &                              &                              & \cellcolor[HTML]{FF8585}1000 & \cellcolor[HTML]{FF9589}840  & \cellcolor[HTML]{FF9589}845  \\
LTC(0.7)  &                              &                              &                              &                              &                              &                              & \cellcolor[HTML]{FF8585}1000 & \cellcolor[HTML]{FF8C87}935  \\
Deg.     &                              &                              &                              &                              &                              &                              &                              & \cellcolor[HTML]{FF8585}1000
\end{tabular}
\end{table*}

\begin{table*}[!h]
\footnotesize\sf
\caption{Overlap in the top 1000 nodes of different centralities on the Pokec network. Measures are ordered according to their distance to Degree. \label{tab:intersection_pokec}}
\begin{tabular}{lllllllll}
           & k-core                       & Harm.                     & TC                           & Shapley(G1)                 &LTC(0.7)                     & PR                     &   GDD(0.05)                 & Deg.                       \\
k-core     & \cellcolor[HTML]{FF8585}1000 & \cellcolor[HTML]{FFE599}42   & \cellcolor[HTML]{FFE599}47   & \cellcolor[HTML]{FFE699}31   & \cellcolor[HTML]{FFE298}80   & \cellcolor[HTML]{FFE399}63   & \cellcolor[HTML]{FFE399}66   & \cellcolor[HTML]{FFDD98}127  \\
Harm.   &                              & \cellcolor[HTML]{FF8585}1000 & \cellcolor[HTML]{FFD396}224  & \cellcolor[HTML]{FFC292}397  & \cellcolor[HTML]{FFAD8E}603  & \cellcolor[HTML]{FFBB90}468  & \cellcolor[HTML]{FFC793}348  & \cellcolor[HTML]{FFBF91}422  \\
TC         &                              &                              & \cellcolor[HTML]{FF8585}1000 & \cellcolor[HTML]{FFCD94}281  & \cellcolor[HTML]{FFCC94}295  & \cellcolor[HTML]{FFC492}373  & \cellcolor[HTML]{FFBC91}460  & \cellcolor[HTML]{FFBC91}453  \\
\multicolumn{2}{l}{Shapley (G1)}          &                              &                              & \cellcolor[HTML]{FF8585}1000 & \cellcolor[HTML]{FFB68F}516  & \cellcolor[HTML]{FFA18B}723  & \cellcolor[HTML]{FFBA90}477  & \cellcolor[HTML]{FFBC91}456  \\
LTC(0.7) &                              &                              &                              &                              & \cellcolor[HTML]{FF8585}1000 & \cellcolor[HTML]{FFA68C}676  & \cellcolor[HTML]{FFB990}485  & \cellcolor[HTML]{FFAD8E}601  \\
PR   &                              &                              &                              &                              &                              & \cellcolor[HTML]{FF8585}1000 & \cellcolor[HTML]{FFA78C}664  & \cellcolor[HTML]{FFA38C}701  \\
GDD(0.05) &                              &                              &                              &                              &                              &                              & \cellcolor[HTML]{FF8585}1000 & \cellcolor[HTML]{FF9D8A}761  \\
Deg.     &                              &                              &                              &                              &                              &                              &                              & \cellcolor[HTML]{FF8585}1000
\end{tabular}
\end{table*}

Table~\ref{tab:avg_med} compiles the average and median date of registration for the top 1000 nodes. Centralities are ordered by the median, last row shows the average and the median for all nodes in the network.

\begin{table*}[t]
\small\sf\centering
\caption{Average and median date of registration of the top 1000 nodes of various centrality measures on the social networks iWiW and Pokec. Registration date is measured in number of days from the kickoff of the network. Last row shows the average/median on the whole network. \label{tab:avg_med}}
\begin{tabular}{llllll}
\hline
 \multicolumn{3}{c}{iWiW}& \multicolumn{3}{c}{Pokec} \\
Centrality  & average   &   median &  Centrality  & average   &   median \\
\hline%
 TC       	&	1554.402	&	1584	&	 TC       	&	2928.189	&	2891.5	\\
 Degree   	&	1657.128	&	1645	&	 GDD(0.05) 	&	3057.238	&	3038	\\
 GDD(0.05) 	&	1648.546	&	1646.5	&	 $k$-core 	&	3145.988	&	3160.5	\\
 LTC(0.7) 	&	1673.296	&	1661	&	 Degree   	&	3169.621	&	3169	\\
 Harmonic 	&	1670.558	&	1662.5	&	 PageRank 	&	3274.957	&	3320	\\
 PageRank 	&	1695.087	&	1669	&	 Harmonic 	&	3300.377	&	3356	\\
 $k$-core 	&	1753.594	&	1704	&	 Shapley (G1) 	&	3357.282	&	3386.5	\\
 Shapley (G1) 	&	1759.33	&	1707	&	 LTC (0.7) 	&	3418.432	&	3435	\\

 \hline
Whole graph	&	1911.242	&	1813 &	Whole graph	&		3163.550	&	3192	
\end{tabular}
\end{table*}

In case of iWiW all measures performed well, that is, all averages/medians are below the network average/median. The Top Candidate (TC) method proved to be the best, with an average date of registration 7\% lower than that of the next best centrality, Degree, and almost 20\% lower than the network average.

TC retains its first place on Pokec as well, though with smaller margins. It performs 4.3\% better than the next best, GDD, and 7.5\% better than the network average. Note that the centralities showed much more volatility: five out of the eight performed worse than the network average.

The results seem to be consistent. TC, GDD and Degree are among the first four, while Harmonic centrality, PageRank and Shapley (G1) lag behind on both networks. Only LTC and $k$-core showed varying results.

The average and median date of registration are, in themselves, imperfect indicators of performance. Due to their extreme risk-aversion, laggards would almost surely refuse to participate in a campaign, while individuals belonging to the early majority might be persuaded with e.g.\ a small financial reward. Hence, we need to take a look at the whole distribution to evaluate the measures.

\begin{figure*}
    \centering
    \includegraphics[width=12cm]{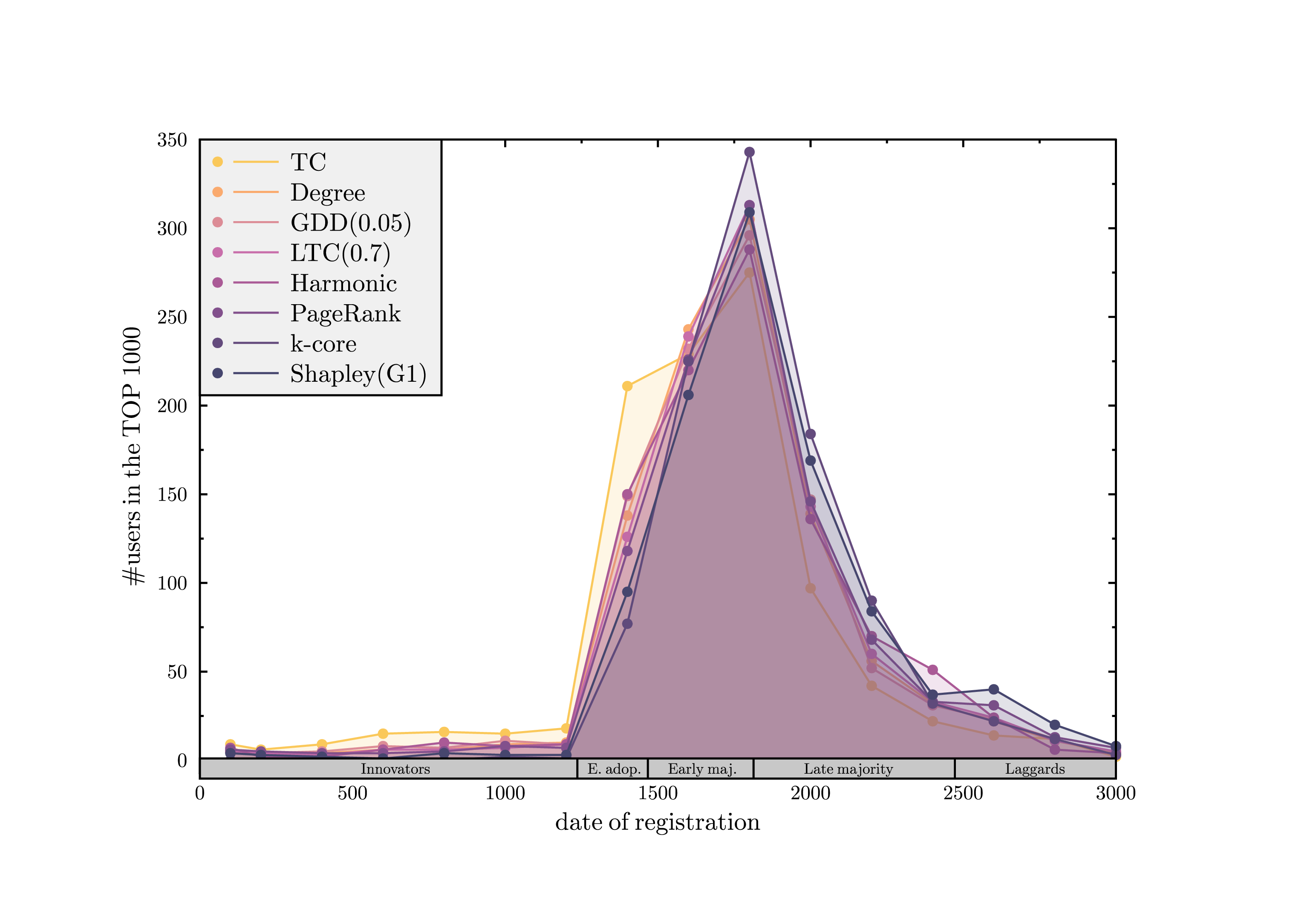}
    \caption{iWiW dates of registration of top 1000 users of various network centralities. Measures are ordered by the median day of registration.}
    \label{fig:iwiw_dist}
\end{figure*}

In case of iWiW the field is mostly even (Figure~\ref{fig:iwiw_dist}). TC is the only centrality that sticks out of the crowd, consistently outperforming the other measures in innovator and early adopter category, while also having the fewest laggards and late majority.

\begin{figure*}
    \centering
    \includegraphics[width=12cm]{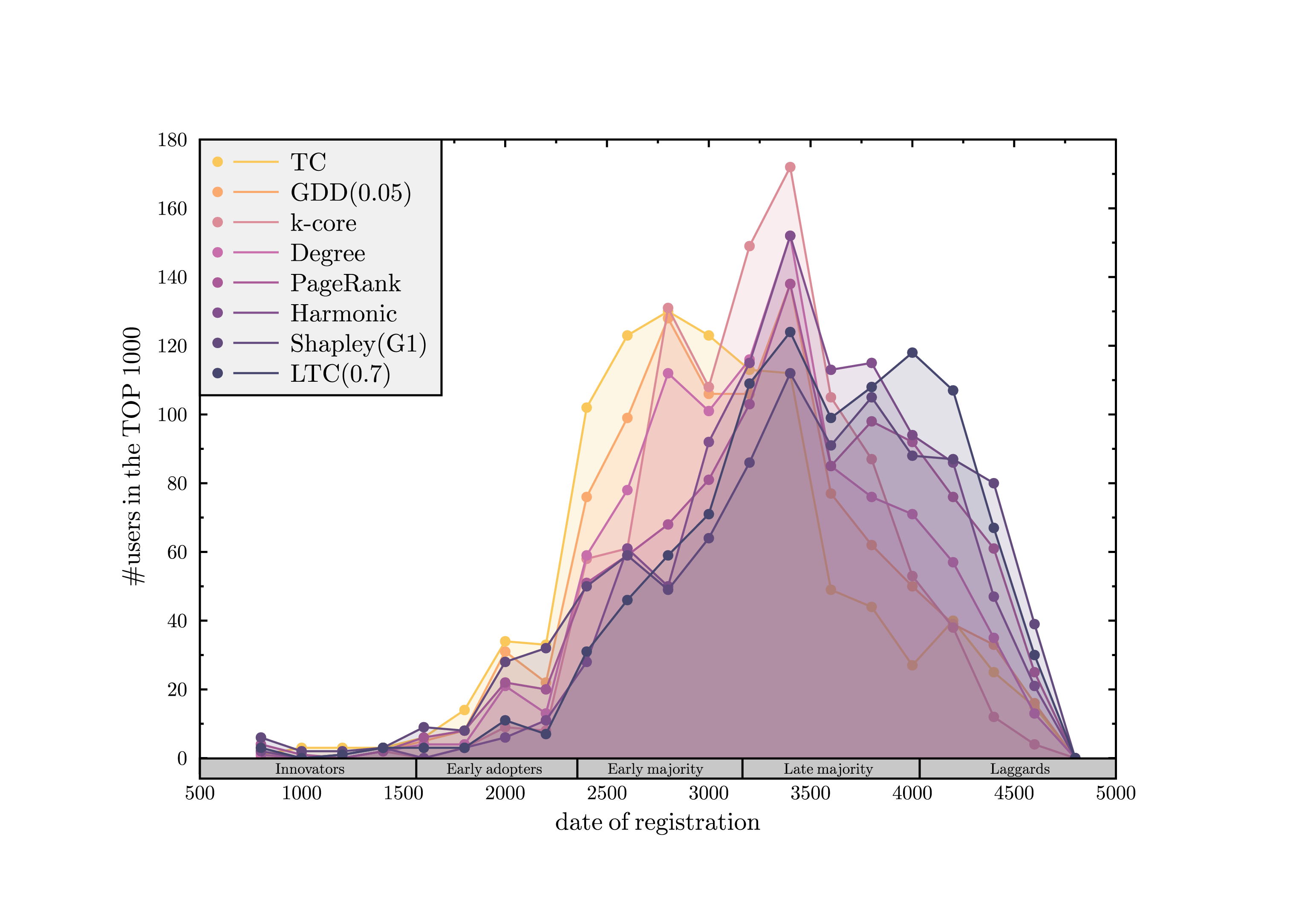}
    \caption{Pokec dates of registration of top 1000 users of various network centralities. Measures are ordered by the median day of registration.}
    \label{fig:Pokec_dist}
\end{figure*}

Although, the performances are more nuanced in Pokec, TC is still the best (Figure~\ref{fig:Pokec_dist}). In case of innovators its performance is on par with the other measures. This is perhaps due to the fact that very few individuals fell into this category. It has more early adopters and early majority and less late majority than any other centrality, while in laggards category it is the second best. GDD also shows some very promising results.

\begin{figure*}
    \centering
    \includegraphics[width=12cm]{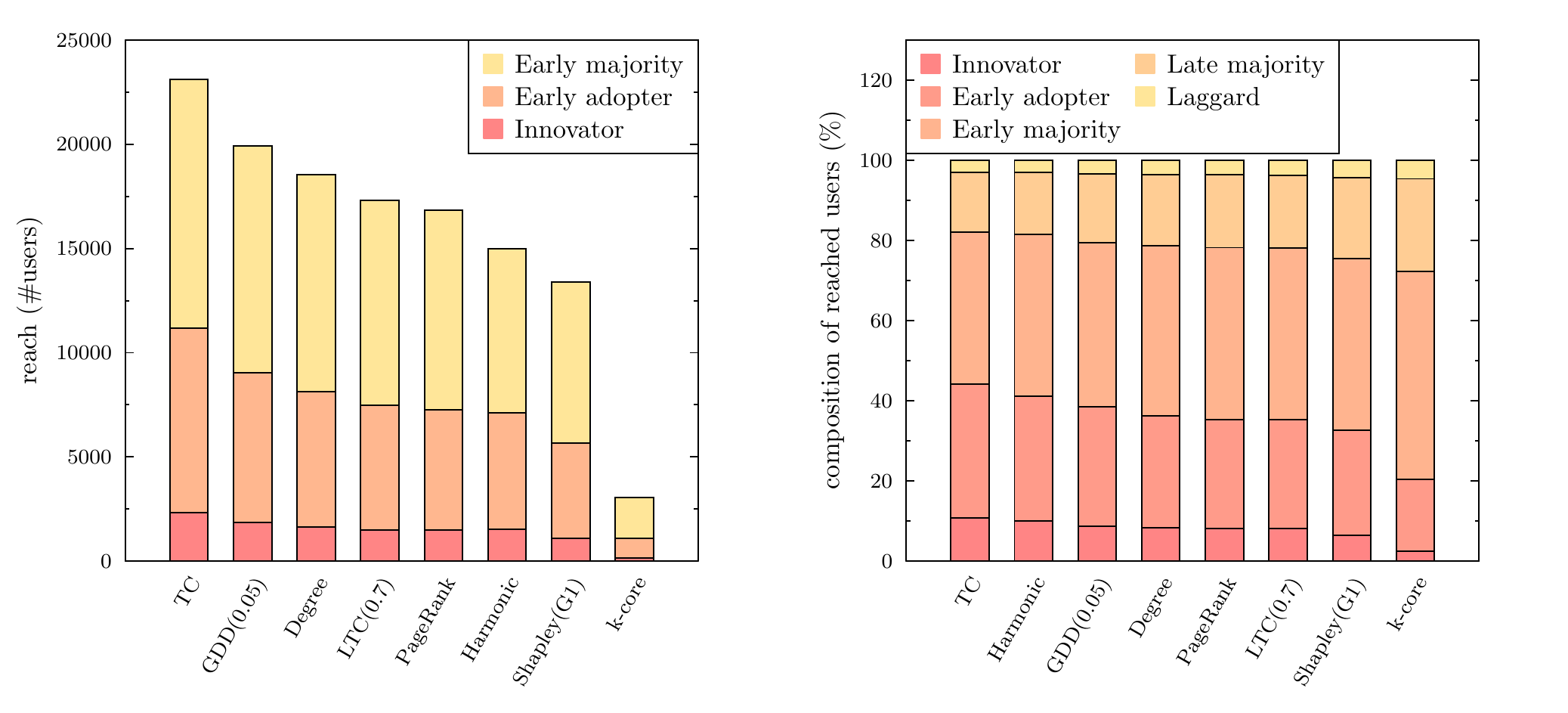}
    \caption{Reach of innovators and early adopters among the top 1000 nodes of different centralities on iWiW.}
    \label{fig:reach_iwiw}
\end{figure*}

\begin{figure*}
    \centering
    \includegraphics[width=12cm]{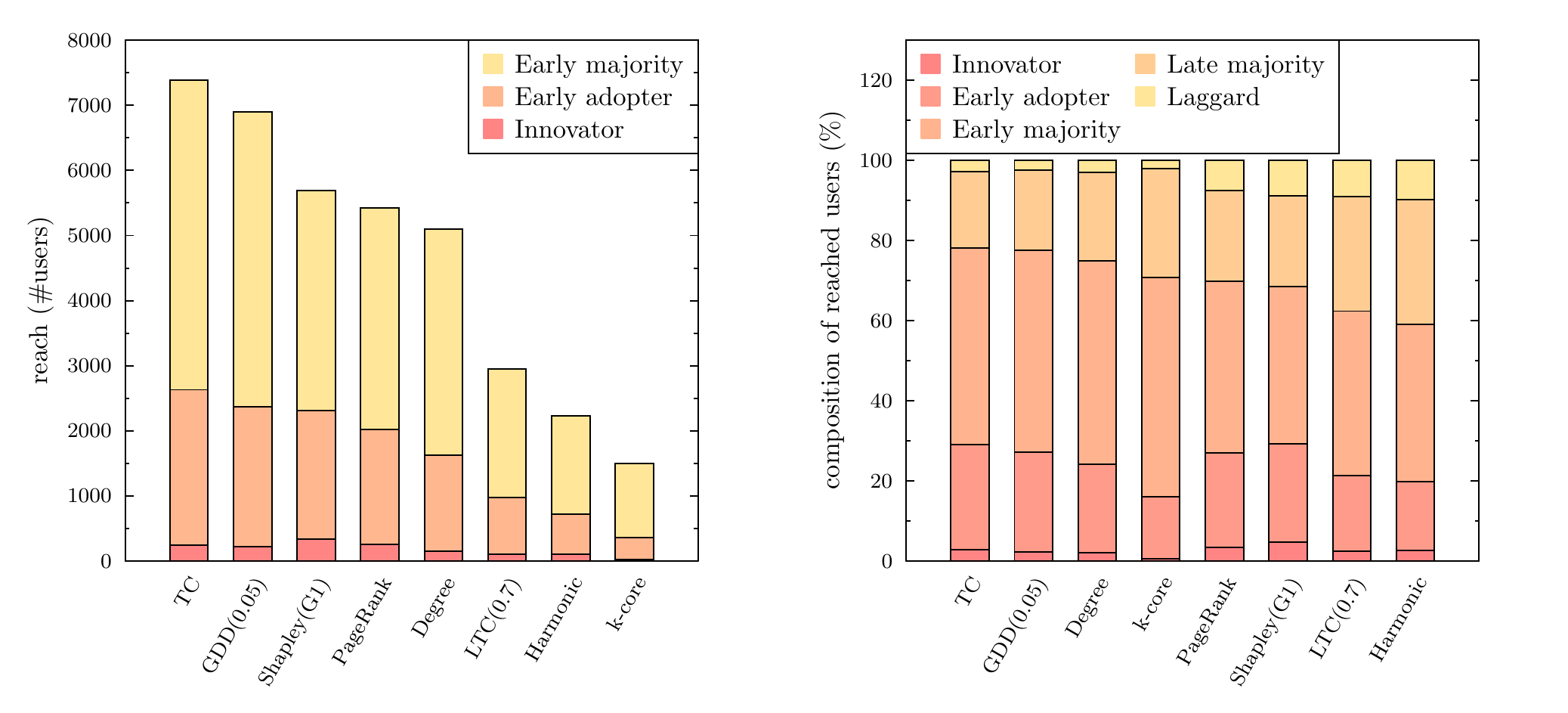}
    \caption{Reach of innovators and early adopters among the top 1000 nodes of different centralities on Pokec.}
    \label{fig:reach_pokec}
\end{figure*}

Assuming that (i) a marketing message or a product sample will only incite innovators or early adopters, and that (ii)  these two groups have their greatest influence on like-minded groups and on early majority,
it is worth to restrict our attention to these two groups and their interactions with their neighbours. Figure~\ref{fig:reach_iwiw} and \ref{fig:reach_pokec} show the net reach of innovators and early adopters among the top 1000. The bar graph on the left depicts how many innovators, early adopters and early majority they reach not counting themselves. This illustrates the indirect impact of the campaign. The bar graphs on the right hand side show the composition of their reach.

Note that TC only comes out as a winner if these two assumptions hold -- the bulk reach  of e.g.~PageRank, that includes late majority and laggards as well, is much larger than that of TC. Thus, on a conventional linear threshold or independent cascade simulation PageRank would outperform TC. However, by omitting these two assumptions we oversimplify the diffusion model and assign inaccurate prediction power to the tested algorithms.

\section{Conclusion}\label{sec:conclusion}

Innovators and early adopters are not abstract theoretical constructions, but groups that can be found on social networks as node clusters with distinct connection preferences. Consequently, they can be identified by observing the network structure. The top choices of some network centralities include more innovators and early adopters than others. Since these two groups play an essential role in innovation spreading, such network centralities might be more effective in real marketing campaigns.

Influence maximization aims to find the most influential nodes on the network. In the past two decades, myriads of clever heuristics were invented to optimize this computationally difficult task. Usually, these algorithms are validated via computer simulations with little care about what a real diffusion would look like. In real life, targeted agents often refuse to participate in the campaign. The underlying reasons are manyfold, but most prominently agents differ in their risk attitudes. No matter how central a node is if it is risk-averse, unwilling to try the advertised product or commit to it openly.

Simulations also commonly ignore network homophily which can have serious impact on how a cascade unfolds. Both social networks presented here show strong patterns of homophily (Table~\ref{tab:gr_inter_iWiW} and \ref{tab:gr_inter_Pokec}).

We tested eight different network centralities on two social networks where data about the date of registration was available. This allowed us to rank the centralities by their ability to identify innovators and early adopters. A novel expert selection algorithm, the Top Candidate method consistently outperformed every other method. To a smaller extent Generalised Degree Discount and Degree were also effective.

A possible explanation of the success of the Top Candidate ranking is that individuals with high socio-economic status and opinion leadership qualities -- two traits that are associated with early adopters -- are perceived as experts in society. Since the Top Candidate method is specifically designed to identify experts, it is a small wonder, that it finds more early adopters than other measures. The Top Candidate ranking is derived by the different parametrizations of the Top Candidate method. For a fixed parameter the Top Candidate method outputs a list of individuals that form a stable set -- the underlying idea is that experts are much more efficient in recognizing each other than amateurs, thus the selected individuals must support each other. This property resembles to assortativity, and might be the reason why the method is successful in identifying such highly assortative sets as innovators and early adopters. Another possible explanation is that TC identifies more market mavens, who are also crucial in innovation spreading and widely acknowledged as experts.



The results may be interesting for practitioners of various fields. Computer scientist often test their heuristics with simulations on either the linear threshold or the independent cascade models. In light of the results, the accuracy of these experiments can be improved by redesigning the threshold and propagation probability distributions. There are already a few papers that study how to obtain sensible propagation probabilities for the independent cascade model but less attention was given to node thresholds, and no papers take into account Roger's adopter classification when calibrating diffusion variables.

For marketing specialists, the practical lessons of this paper is that they should aim for experts in a campaign, and that the Top Candidate method is an excellent tool for finding them.



\section*{Acknowledgments}

The authors acknowledge financial help received from National Research, Development and Innovation Office, grant numbers K 138945 (Sziklai) and KH 130502 (Lengyel). Bal\'azs R.\ Sziklai is the grantee of the J\'anos Bolyai Research Scholarship of the Hungarian Academy of Sciences. Supported by the ÚNKP-22-5 New National Excellence Program of the Ministry for Culture and Innovation from the source of the National Research, Development and Innovation Fund.

\bibliography{Innovators}

\end{document}